\documentclass[apj]{emulateapj}

\newcommand{\etal}{\emph{et\,al.}}
\newcommand{\flames}{\textsc{flames}}
\newcommand{\giraffe}{\textsc{giraffe}}
\newcommand{\argus}{\textsc{argus}}
\newcommand{\idl}{\textsc{idl}}
\newcommand{\iraf}{\textsc{iraf}}
\newcommand{\rrb}{$\lambda$5800\AA}
\newcommand{\dib}{$\lambda$5797\AA}

\newcommand{\hd}{\objectname[HD44179]{HD\,44179}}
\newcommand{\rr}{\objectname[red rectangle]{Red Rectangle}}
\newcommand{\vlt}{\facility[VLT]{\textsc{vlt}}}
\shorttitle{Sequence Structure emission in The Red Rectangle}
\shortauthors{Sharp, Reilly, Kable and Schmidt}

\begin{document}

\title{Sequence structure emission in The Red Rectangle Bands}

\author{R.G. Sharp}
\affil{Anglo-Australian Observatory, PO Box 296, Epping, NSW, 1710, Australia}
\email{rgs@aao.gov.au}
\and
\author{N.J. Reilly, S.H. Kable, T.W. Schmidt}
\email{T.Schmidt@chem.usyd.edu.au }
\affil{ School of Chemistry, Building F11, University of Sydney, NSW,
2006, Australia}

\begin{abstract}
We report high resolution (R$\sim$37,000) integral field spectroscopy
of the central region (r$<$14\,arcsec) of the \rr\ nebula surrounding
\hd.  The observations focus on the \rrb\ emission feature, the bluest
of the yellow/red emission bands in the Red Rectangle.  We propose
that the emission feature, widely believed to be a molecular emission
band, is not a molecular rotation contour, but a vibrational contour
caused by overlapping sequence bands from a molecule with an extended
chromophore. We model the feature as arising in a Polycyclic Aromatic
Hydrocarbon (PAH) with 45-100 carbon atoms.
\end{abstract}


\section{Introduction}
The Diffuse Interstellar Bands (DIBs), first identified as
interstellar absorption features by Merrill and coworkers
\citep{Merrill1934,Merrill1936,Merrill1938}, represent one of the
longest outstanding mysteries in astronomical spectroscopy.

The DIBs are widely believed to represent the origin bands of
molecular transitions.  Ultra-high resolution spectroscopy of a
selection of the stronger DIBs confirms their molecular origin
\citep{Sarre1995a,Kerr1998}, while the lack of strong correlations
between known bands precludes the identification of any
transitions to higher vibronic states.

Identification of the carriers of the DIBs will ultimately require
accurate measurement of the gas phase spectrum of a range of
extraterrestrial molecules to allow direct comparison with
astronomical observations (for a review of current work see
\citet{Schmidt2005,Fulara2000,Herbig1995}). However,
the parameter space occupied by the range of candidate carriers is
prohibitively large for full investigation with the often complex
techniques required to record the gas phase spectra.

In order to restrict the parameter space one can turn to studies of
alternate environments to the diffuse clouds within which most DIBs
are observed. These alternate environments may represent the
\emph{Rosetta stone} for DIB spectroscopy\footnote{We thank A. Witt
for suggesting this elegant metaphor, private communication},
offering the possibility of presenting additional features of the
molecular species, providing the much needed additional constraints to
the properties of the carrier molecules.

In this paper we present a high spatial, and spectral, resolution
study of one of the emission features seen in the \rr\ nebula.  At a
distance of only 330pc (but see \citet{Men'shchikov2002} for details
of a revised distance estimate of 710pc\footnote{Hipparchus satellite
measurements give a value of 2.62$\pm$2.37mas/yr corresponding to the
range $\sim$200-4000pc.}) the nebula represents an excellent candidate
for the study of the injection of material into the interstellar
medium by a dying star. 

\section{The Red Rectangle}
The bi-conical \rr\ (RR) nebula \citep{Cohen1975,Cohen2004}
surrounding the evolved, mass-losing star \hd\ is one of the strongest
sources in the sky when observed in the mid-IR (IRAS\,06176-1036),
where a set of emission features commonly attributed to polycyclic
aromatic hydrocarbons (PAHs, \citet{Allamandola1985}) dominate the
spectrum.  PAH molecules provide an elegant solution to a number of
astronomical issues ( UV exctinction curve - \citep{Donn1968}; DIBs
carriers - \citep{Zwet1985,Leger1985}; UIB
carriers\citep{Geballe1989}).  With the coming of age of the new
generation of mid IR instrumentation, PAH emission may prove to be a
valuable tracer of extragalactic star formation \citep{Peeters2004}.
However, at this time, not a single PAH has been unambiguously
identified in the ISM (but see \citet{Allamandola1999} for details of
composite spectra of the unidentified Infra-red Bands (UIBs) in the
mid-IR).  At the heart of the problem is the simple fact that while
the $\lambda$$\lambda$3.3, 6.2, 7.7, 8.6 \& 11.3$\mu$m emission
features (\citet{Peeters2004}) most likely represent excellent tracers
of PAH emission, their origins in the C-H and C-C stretches, abundant
in PAH molecules, mean they present little or no diagnostic power.  If
one is to use PAH emission as a key diagnostic on interstellar
processes, within and outside of our Galaxy, it would seem prudent to
understand the phenomenon.

Spatially extended optical emission bands near \rrb\ were first
reported by \citet{Schmidt1980} but currently remain unidentified.  A
number of detailed descriptions of the form of the \rr\ Bands (RRB)
are available in the literature.  We refer the reader to
\citet{VanWinckel2002}, rather than repeat a detailed description.
Following \citet{VanWinckel2002} we adopt the standard molecular
terminology \emph{Red Degraded} (RD) to refer to RRBs which show a
sharp edge in the blue but an extended tail to the red.  The \rrb\
band is a typical RD RRB.  Several authors have noted the close
correspondence of many of the RRBs with prominent DIBs
\citep{Scarrott1992, Sarre1995b, VanWinckel2002}.  While initial
observations found strong support for convergence of the RRBs to DIBs,
the early works were hampered by low resolution and low
signal-to-noise observations. Later works \citep{Glinski2002} indicate
that the blue-ward shifts of the RRBs, with increasing nebular radius,
do not appear to asymptotically approach the DIB wavelengths.  We
concur that the \rrb\ RRB, in emission does not blue shift to the
wavelength of the prominent \dib\ DIB, at least out to a nebular
radius in excess of 14.0arcsec. \citet{Glinski1997} initially proposed
C$_{3}$ as the carrier of the \rrb\ \rr\ bands.  However, more recent
observations suggest this is not the case
\citep{Glinski2002}. Molecular band structures are reported towards
the centre of the nebula with \citet{Miyata2004} presenting molecular
PAH emission in the inner regions of the \rr\ in the \emph{N} band at
$\sim$10$\mu$m.

The observations of \citet{Glinski2002}, from the Densepak IFU on the
WIYN telescope, show radial evolution of the molecular emission
features.  However, the low surface brightness\footnote{\emph{HST} PC
F622W m$<$17.6mag(Vega) arcsec$^2$} of the \rr\ in the outer regions
of the nebula probed by \citet{Glinski2002} (10-14arcsec) has hampered
previous attempts to identify the molecular spectra.

Recent years have seen considerable progress in laboratory
determination of the spectra of complex astrophysically relevant
molecules \citep{Thaddeus2001,Linnartz2004,Schmidt2003a,Ding2003}.
There has also been progress in the astronomical detection of
molecular species in the absorption spectra of diffuse clouds
\citep{Maier2001,Adamkovics2003} and in millimetre-wave emission.
However, the low surface brightness and complex spatial structure of
the \rr\ have prevented the identification of the carrier of its
strong molecular emission.

\section{Observations and data reduction}
We used the \flames\ instrument at one of the Nasmyth foci of Unit
Telescope 2 (UT2 - Kueyen) of the European Southern Observatory's Very
Large Telescope (ESO \vlt) located upon Cerro Paranal in northern
Chile.  The observations were undertaken in visitor mode during
29$^{\rm{th}}$ and 30$^{\rm{th}}$ December 2004 as part of the
Australian guaranteed time awarded for the construction of the OzPoz
Fibre positioner by the Anglo-Australian Observatory
\citep{Gillingham2003}. The \flames\ instrument comprises a robotic
fibre positioner which provides a range of fibre optic feeds to the
bench mounted \giraffe\ spectrograph.  We used the \argus\ Integral
Field Unit (IFU) to record spectra of a 2D region of the nebulosity.
The \argus\ IFU, a rectangular array of 22$\times$14 micro lenses, can
provide two spatial sampling scales, 0.52arcsec and 0.3arcsec,
yielding fields of view of 11.44$\times$7.28arcsec and
6.6$\times$4.2arcsec respectively.  We used the 0.52arcsec lens scale
for our observations due to the low surface brightness of the nebula
in the regions of primary interest.

\flames\ is operated in a series of standard user modes.  We use the
HR11 mode for our observations, with a central wavelength of 5728\AA\
and covering a wavelength range of 5597-5840\AA\ at a dispersion of
$\sim$0.06\AA/pixel and a resolution element of 2.6pixels (0.156\AA,
R$\sim$37,000). The detector was a 2k$\times$4k EEV CCD operated in
the standard \flames\ readout mode.

\begin{figure}
\plotone{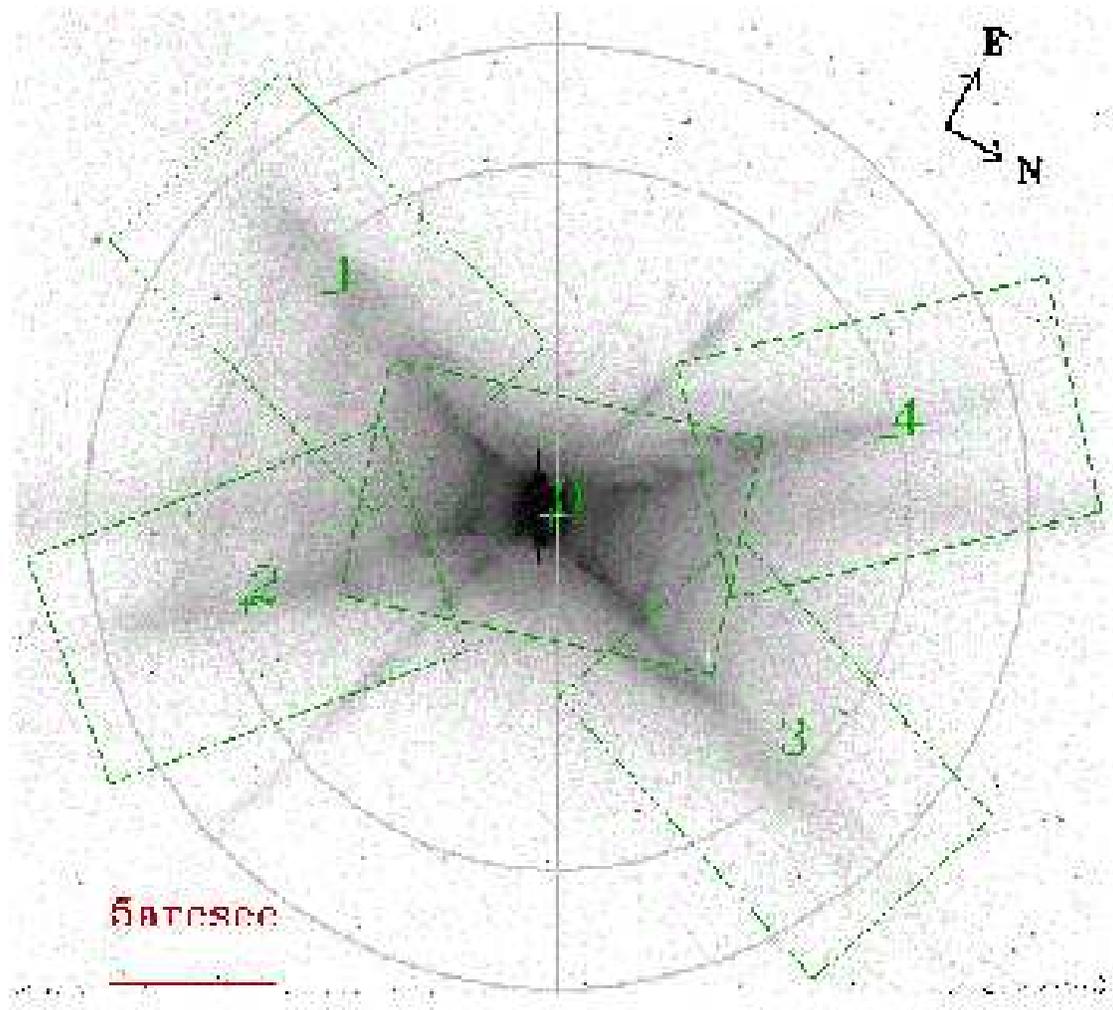}
  \caption{\label{IFU locatation} The locations and orientations of
  base telescope pointings and the \argus\ IFU are shown overlayed on
  the \emph{HST} PC image, at F622W, of \citet{Cohen2004}.  The large
  circles are drawn at a radius of 10.7 and 14.3arcsec, which
  correspond to the fiducial WIYN Densepak spectra of
  \citet{Glinski2002}.}
\end{figure}

Table \ref{observations} and Figure \ref{IFU locatation} detail the
observations undertaken.  Observations were performed during two
nights in classically scheduled time, 29$^{\rm{th}}$ and
30$^{\rm{th}}$ December 2004.  While classical mode was not ideal for
the observations, due to the large range in airmass for the target
during the night (zenith distance
60$^{\circ}$-10$^{\circ}$-60$^{\circ}$), classical mode is a
requirement for guaranteed time observations.

\argus\ attached flat frames are observed, after standard star
spectra, twice per night, in order to provide relative fibre
transmission information and fibre position traces.

Wavelength calibration is derived from standard \vlt\ \flames\ daytime
calibrations.  The \giraffe\ spectrograph offers the ability to
observe a number of simultaneous calibration arcs from a Th-Ar lamp
during the science observations.  Unfortunately, a strong line is
located close to the primary wavelength region of interest leading to
significant contamination of the low light level nebula spectrum
during a 600sec exposures and so simultaneous calibration was not
used.

After discussion with ESO support staff we used the \argus\ fast
acquisition method whereby acquisition relies on accurate coordinates
for a guide star and will be accurate to the order of 1-2arcsec.  A
more accurate, and repeatable acquisition would incur a considerable
overhead for repositioning of guide fibre bundles during the night
(with each change in PA).  The loss of positional accuracy has not
degraded the observations presented.

Five base positions were observed within the nebula.  One centred on
\hd, and four aligned along the arms of the nebula, where limb
brightening of the bi-conical outflow gives the nebula its highest
surface brightness for a given radial distance, and extending out to
$\sim$14arcsec.  Figure \ref{IFU locatation} illustrates the IFU
locations within the nebula.  A three point telescope dither was
performed at each location to allow removal of IFU lens artifacts and
detector defects, of which there are few in the \argus/\giraffe\
system.

\argus\ sky fibres were placed, as required by the \argus\ fast
acquisition strategy, in a ring centred on the IFU.  The ring radius
was set to 3arcmins after visual inspection of the DSS image of the
region showed that this would avoid the majority of brighter stars in
the field regardless of position angle.  This avoids the need to
reposition the fibres between observations which would introduce a
significant overhead.

Simultaneous observations with the UVES fibre system were not possible
due to limits on the proximity of UVES fibre placements to the \argus\
IFU.

\subsection{Data reduction}

During the observing period, pipeline data processing was not
available for the HR11 \argus\ setting used.  We therefore processed
the data using a suite of custom written \idl\ routines and elements
of \iraf.  In most cases, custom \idl\ implementations of common
\iraf\ tasks are used to allow accurate propagation of error
information.

A 2D bias subtraction was not deemed necessary on inspection of the
data and hence the data were overscan corrected only.  On preliminary
data reduction it was decided a known dark current feature seen on the
upper left corner of the \giraffe\ CCD represented a significant
contamination to the low light levels recorded in the outer reaches of
the nebula and hence subtraction of a dark frame was required.  A
master dark was constructed from 6$\times$1200sec dark exposures.
This master dark was scaled for the 600sec and 1800sec exposures as no
alternate dark frames were available.

Flat fields are combined using variance weighting with outlier
rejection.  Where three or more science observations are available, at
a constant dither position, these frames are also combined in a
similar manner.

A fibre trace is generated using the \iraf\ apall task, applied to the
flat field frame.  The database trace output from the task (a series
of 5$^{\rm{th}}$ order Legendre polynomials) was then used to provide
the fibre centres for an optimal extraction algorithm which accounts
for fibre cross talk in the spectra.  Crosstalk was perceived to be a
problem during the initial data reduction due to the high surface
brightness contrast between the inner and outer nebula.

The extracted fibre flat frames are applied to the science frames
after the spectra are extracted in order to retain the projected
relative intensity information between the fibres for the extraction
process.

A wavelength solution was derived, using the \iraf\ identify and
reidentify tasks (fitting a 6$^{\rm{th}}$ order Legendre polynomial,
with a RMS residual of the order of 0.01\AA, for each fibre) using
ThAr arc exposures.  The spectra are then independently transformed to
a common wavelength solution.  A single 2D transformation, as might be
used for long slit data, is not possible due to discontinuities in the
wavelength solution as a function of fibre number at the boundaries
between the sub slits which make up the \argus\ slit in the \giraffe\
spectrograph.

On rectification an estimate of the flat field lamp illumination
spectrum is created from the rectified flat field and multiplied back
into the spectra to remove the lamp signature.

An absolute flux calibration for the data is not required.  However, a
good relative calibration internal to the spectrum is important for
molecular modeling.  A response function is derived via repeated
observations of the standard star EG-21, a DA white dwarf with few
spectral features in the wavelength range of interest and favourably
located on the night sky during observations.

Spectra are subsequently corrected to a heliocentric wavelength
solution.

\subsection{Mosaicing the data}
Data from individual cosmic-ray splits (typically 3$\times$1200sec)
are combined using variance weighting and outlier rejection.  Three
dither positions were observed at each location in the nebula.  In
order to align the dither data, and to facilitate the ultimate
combination of all five telescope pointings into a uniform mosaic, we
re-sample the data onto a common regular Cartesian grid after applying
the appropriate position angle (PA) rotation to the IFU internal x/y
coordinate system, retrieved from the \flames\ header, to account for
the five different position angles used during the observations.  A
number of techniques were subsequently investigated for aligning
individual data cubes.  Cross-correlation of pseudo-continuum images
created from the data cubes proved difficult, due to small image
overlaps.  Ultimately alignment was performed using an iterative
$\chi^2$ procedure, and the \idl\ implementation of the Amoeba
algorithm \citep{numrepc}, as applied to a subsection of each
dispersed data cube.

The final analysis of the full data set is ongoing.  Data from only a
single arm of the nebula (pointing 2 in Figure \ref{IFU locatation})
is used to illustrate our model for the \rrb\ emission.  Preliminary
investigations show that the emission in the remaining three nebula
arms are qualitatively similar.

\section[]{Interpreting the $\lambda$5800\AA\ band.}
A number of origins have been proposed for the \rrb\ emission.  Most
interpretations propose the band profile represents a molecular
rotation contour \citep{Scarrott1992, Rouan1997, Glinski1997}.
\citet{Glinski2002} demonstrate that the \rrb\ cannot be generated via
Q, R or P branch rotational emission structures derived from the
unknown molecular carrier DIB at \dib.  \citet{Rouan1997} propose that
the RRB has its origins in a supra-thermally rotating PAH molecule,
and go on to argue that a PAH of around 40 carbon atoms is implied by
the required rotational constants. This model assumes that the DIB
carrier at \dib\ is identical to the \rrb\ RRB carrier and that
angular momentum is accumulated due to a ``rocket effect'' induced by
photo-dissociation events. This effect is required to attain an
estimated rotational temperature of 450\,K for the \rrb\ carriers.
Rotational band profiles are always expected of molecular spectra
since there exist a large number of energetically accessible
rotational states, even when the molecule in question is
electronically and vibrationally cold. Thus a single vibronic
transition will be derived from a number of individual transitions
originating in different rotational states. Energetically accessible
vibrational states, potentially pumped by photoexcitation, can produce
hot-band features in an electronic spectrum, where the vibronic
transitions originate from one of these excited vibrational states.
Where the potential energy surfaces (PES), governing the forces on the
nuclei in the electronic states, are parallel, the structure of the
excitation or emission spectrum will be dominated by sequence bands.
This will be the case for a molecule with an extended chromophore
since the shape of the underlying molecule will be largely unaltered
upon electronic excitation.

\subsection[]{Sequence structure}
\begin{figure}
\plottwo{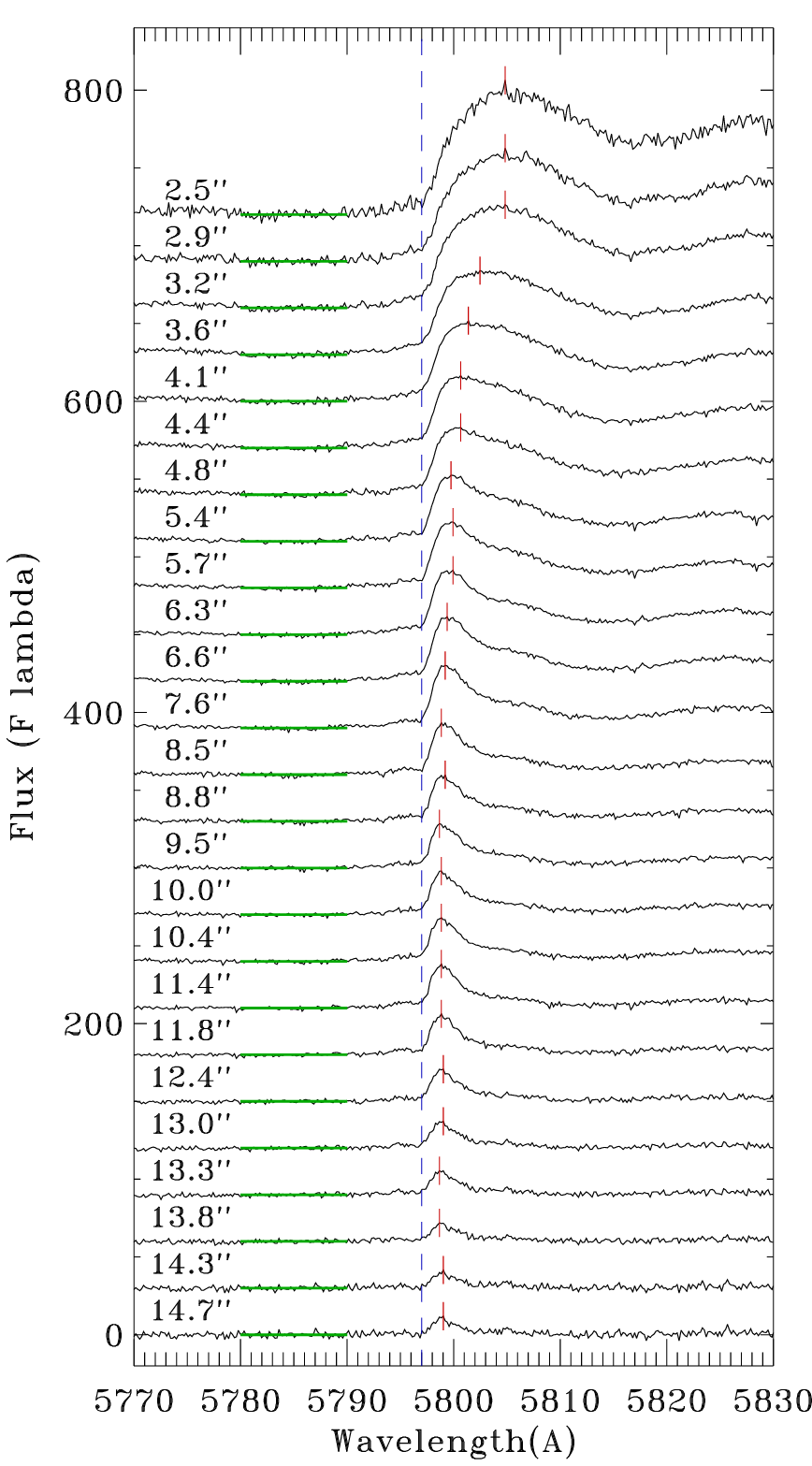}{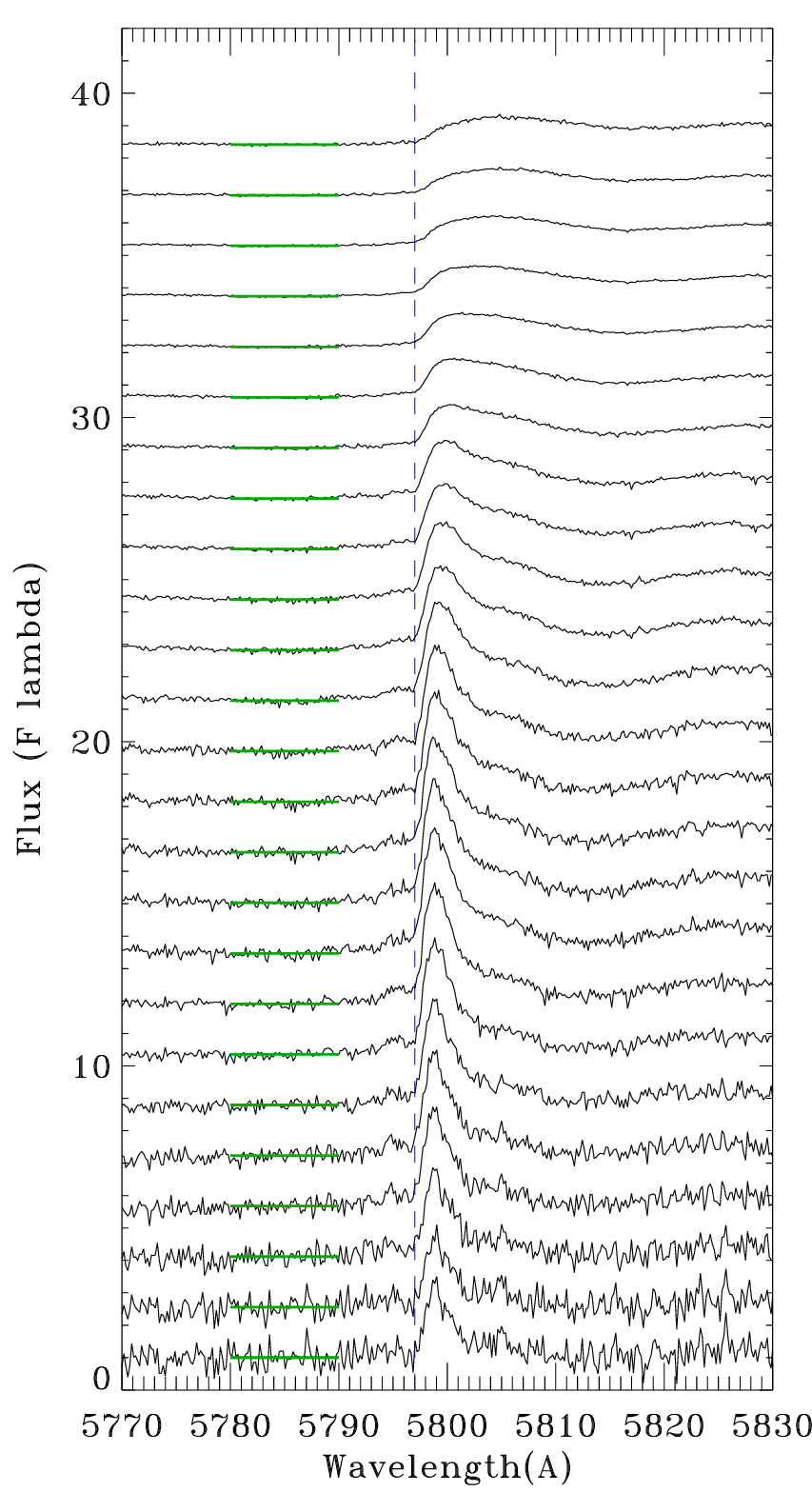}
  \caption{\label{sequence-arm2} As an example of the strong evolution
  of the \rrb\ band, as a function of radial distance from the central
  star, a sequence of spectra from the south western arm (P2 in Figure
  \ref{IFU locatation}) of the nebula is shown.  The radial distance
  from HD\,44179 is indicated to the left.  The spectra are vertically
  baseline offset for clarity.  Spectra in the left-hand Figure have
  been continuum subtracted, while the identical spectra in the
  right-hand Figure have been normalised by the continuum. In this
  illustration the continuum has been estimated as a simple constant
  value over the spectral range $\lambda\lambda$5780-5790\AA .  A
  vertical dashed line illustrates the location of the
  $\lambda$5797\AA\ DIB.  Tick marks indicate the \emph{peak} of the
  \rrb\ band within each spectrum.}
\end{figure}

Based on our study of the radial evolution of the emission structure
along the arms of the bi-cone (Figures \ref{IFU locatation} \&
\ref{sequence-arm2}) we propose that the \rrb\ feature arises from
sequence structure, either associated with the nearby \dib\ DIB
carrier, or an alternate molecule.  The absence of any perfect
correlations between DIBs leads one to conclude that if the DIB
carriers are molecular \citep{Thorburn2003} then the Franck-Condon
factors for electronic excitation are dominated by the origin
band. For a molecule in its vibrational and electronic ground state,
electronic excitation will similarly leave the molecule with zero
vibration in the electronic excited state.  The dominance of the
origin band in electronic excitation is indicative of a transition
between states with very similar potential energy surfaces. If the
equilibrium geometry or the vibrational frequencies were to change
appreciably upon excitation, vibrational progressions would be
observed.

If the carrier of the \rrb\ feature behaves similarly to the DIBs,
then we would expect any emission from a vibrationally excited
population of molecules to exhibit sequence structure. Sequence
structure arises from electronic transitions where the molecule
remains in vibrational states of the same character. If the potential
energy surfaces of two electronic states are parallel, then the
sequence emission will occur at the same wavelength as the origin
band.

In many large molecules, electronic transitions between states of
increasing vibrational excitation result in transitions of
progressively lower energy due to the tendency for vibrational
frequencies to be \emph{slightly} lower in the excited electronic
state. In large molecules where electronic transitions from the ground
state are dominated by the origin transition, population of higher
vibrational excitation states results in a red-ward broadening of the
emission structure.

This sequence structure interpretation requires the molecule
responsible for the emission to have an extended chromophore, in order
that the electronic transitions do not change the structure of the
molecule appreciably (\emph{i.e.} the PESs are almost parallel).

In deference to the current popularity in astronomy of PAHs we, in the
first instance, follow the trend and propose an intermediate size PAH
molecule as the emitter. Alternative molecular forms (C$_{\rm{60}}$ -
\citet{Kroto1985}, carbon chains - \citet{Douglas1977}), are not
considered at this time.

Using the Ultra-High Resolution Facility (UHRF, R$\sim$600,000) of the
Anglo-Australian telescope (AAT), \citet{Sarre1995a} show that the
\dib\ DIB carrier fulfils the requirements to generate sequence
structure. The slightly red-degraded shape of the UHRF spectrum shows
that the molecule is slightly larger in the excited state, indicating
a weakening of bonds and thus a probable slight decrease in the
vibrational frequencies of the molecule. It is thus suggested that if
a vibrationally excited population of the carrier of the \dib\ DIB
were to fluoresce (or phosphoresce) it would exhibit sequence
structure as manifested by red-ward broadening of the DIB band
shape. However, as the \rrb\ feature does not asymptotically approach
the \dib\ DIB position, the simplest assumption to be made is that
they arise from a different carrier. For the bands to arise from the
same carrier we require that the coldest emission be completely
self-absorbed by the nebula.

\subsection[]{PAH Model}
In order to explore the sequence band hypothesis, we performed
normal mode analyses on PAH cations of various sizes in ground and
electronically excited states. Cations, rather than neutrals, were
chosen for these quantum calculations so that they exhibited strong
transitions in the visible region \citep{Brechignac1999}. Ground
state geometries and normal mode frequencies of the naphthalene,
pyrene, tetracene, perylene and coronene cations were obtained with
the B3LYP density functional and 6-31G basis set using the
\textsc{gaussian 03} \citep{Frisch2003} suite of programs.
Time-dependent Density Functional Theory (DFT) calculations were
performed on the optimised ground state structures to yield excited
states. The electronic transition in the visible region with the
largest oscillator strength was chosen. In order to replicate this
excited state, the ground state orbital occupancy was altered to
resemble the leading configuration of the selected excited state.
Geometry optimisations and normal mode analyses were performed on
these excited states for naphthalene, pyrene, tetracene and
perylene.

The vibrational frequencies in the ground $\{\nu_i''\}$ and excited
states $\{\nu_i'\}$ were used to generate transition energies with the
origin transition (arbitrarily) centred at $\lambda$5800\AA, in order
that we be able to compare the band shape with the new RRB
observations. At the time of writing, only indicative origin positions
may be calculated by quantum theory for molecules of the size being
invoked here. By way of benchmarking, pentacene is seen to absorb at
$\lambda$5362\AA\ \citep{Heinecke1998} but is calculated by TD-DFT to
absorb at 6190\AA\ (Reilly \& Schmidt, private
communication). Sequence band profiles were calculated in the manner
laid out in appendix \ref{dead hard sums}.

For an example of an observed molecular sequence structure see Figure
2 of \citet{Linnartz2004}.

\subsection[]{DFT Modeled emission spectra}
It must be emphasised that the modelling of emission spectra is, in
essence, an exploration of the parameter space available to the
molecules responsible for the \rrb\ emission. As such, by invoking
naphthalene, pyrene, tetracene and perylene cations we are exploring
the expected vibrational sequence structure of a molecule of that
size and shape, with an extended chromophore, rather than the
species itself.

\begin{figure}
\plotone{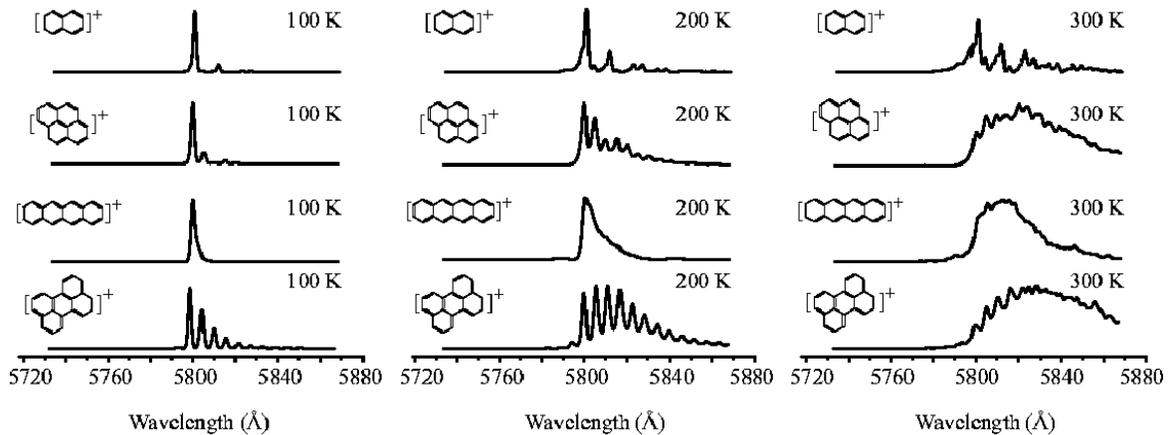}
  \caption{\label{small PAH} Calculating the sequence structure
  emission profile using Density Functional Theory (DFT), we conclude
  that small PAHs alter their structure considerably on excitation.
  Consequently, the red-shift in the band peak is too great and too
  much structure is resolved within the band profile for molecules of
  this size to represent the \rrb\ band evolution.}
\end{figure}

Notwithstanding the fact that naphthalene cation is likely
photodestroyed in harsh circumstellar environments, it serves as a
starting point for the exploration of our parameter space. As can be
seen in figure \ref{small PAH}, the modeled naphthalene spectra are
far too structured and cover too large a wavelength region to be
responsible for the RRBs. What's more, the density of states does
not increase fast enough with energy to bring about a shift in the
maximum of the emission spectrum.

On moving to pyrene, we see that the spectra become less resolved due
to an increased number of emissive states. At first glance it would
appear that pyrene behaves in a very similar manner to the new RR
spectra. However, the shift in the emission peak for pyrene is too
large to be considered a serious candidate for the RR emission carrier
(\emph{cf.} Figures \ref{sequence-arm2} \& \ref{small PAH}).

Tetracene is calculated to have small vibrational frequency shifts
as compared to naphthalene and pyrene. As such it yields narrow
sequence structure features which are broadly in keeping with the
behaviour observed in the RR.

Perylene behaves in a similar way to pyrene yet its sequence
structure is calculated to be somewhat dominated by a small
frequency mode with a relatively large shift upon excitation. The
greater number of Boltzmann accessible vibrational states accessed
in the electronic excited state brings about a shift in the maximum
of the emission spectrum to longer wavelengths. However, even for
perylene the magnitude of the red-shift of the calculated sequence
structure is far greater than that observed in the RRBs. In order
that the RRBs be explained by sequence structure, we require a
larger molecule with a much larger density of states.

The size of the molecule required is such that the modeling of its
sequence structure explicitly using DFT becomes intractable. While
it may be possible to obtain ground state electronic structures and
vibrational frequencies of very large molecules, the excited state
electronic structures become increasingly complicated, being brought
about by mixing several excitations out of the ground state. As
such, representation of an excited state with an altered orbital
occupancy is unreliable. Optimizing excited states by using TD-DFT
is beyond the scope of contemporary computer codes for the size of
molecules being employed here (\emph{vide infra}).

\begin{figure}
\epsscale{.8}\plotone{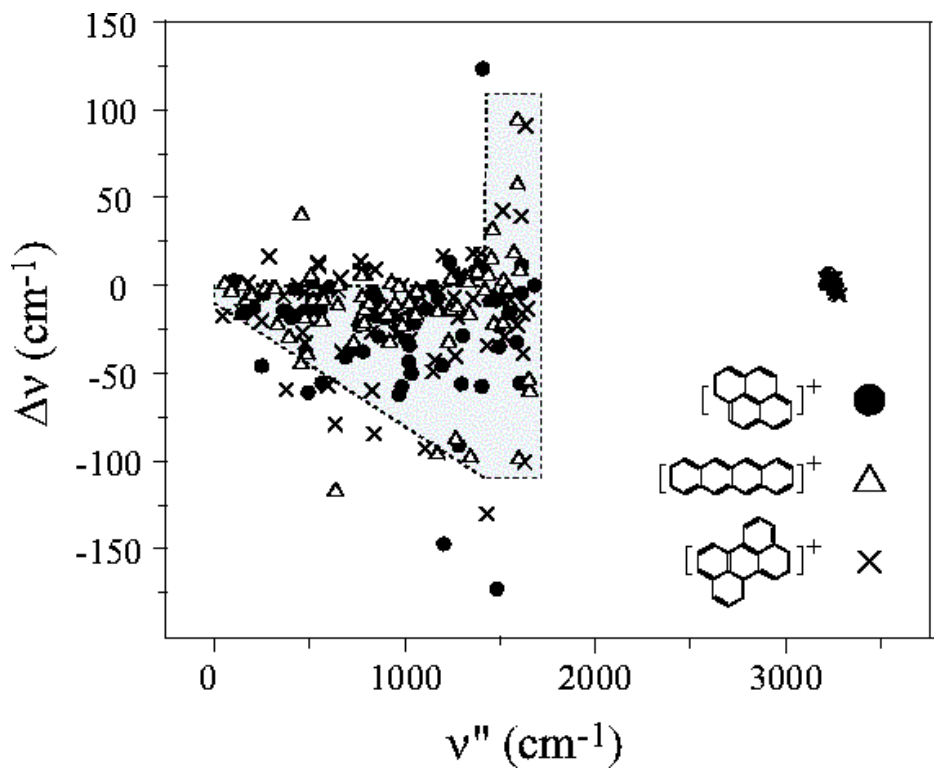}\epsscale{1}
  \caption{\label{shifts}A plot of ground state DFT vibrational
  frequencies ($\nu''$) against their respective shifts in the excited
  electronic state ($\Delta\nu$).  For molecules larger than perylene
  explicit DFT calculations for the ground and excited states are
  intractable. In order to model sequence structure emission for
  larger molecules, to demonstrate how such emission would be
  observed, we assume a series of randomly realised frequency shifts
  drawn from the envelope of parameter space populated by a number of
  PAH molecules for which explicit DFT calculations of both the ground
  and excited states were possible.}
\end{figure}

\begin{figure}
\epsscale{.8}\plotone{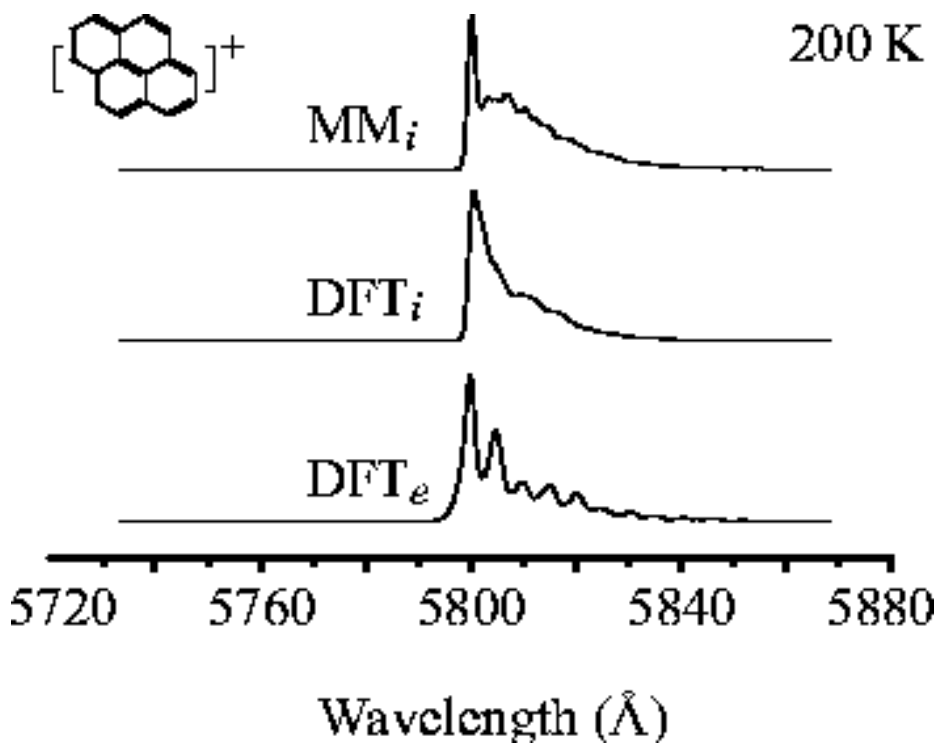}\epsscale{1}
  \caption{\label{dovetail} For the PAH pyrene explicit DFT
  calculations of the ground and excited state spectra are possible.
  Here we contrast the \emph{explicit} DFT calculations (DFT$_{e}$)
  with the \emph{implicit} model outlined in the text using the ground
  states frequencies derived from the DFT calculations (DFT$_i$) and
  the MM model (MM$_i$).}
\end{figure}

\begin{figure}
\epsscale{.8}\plotone{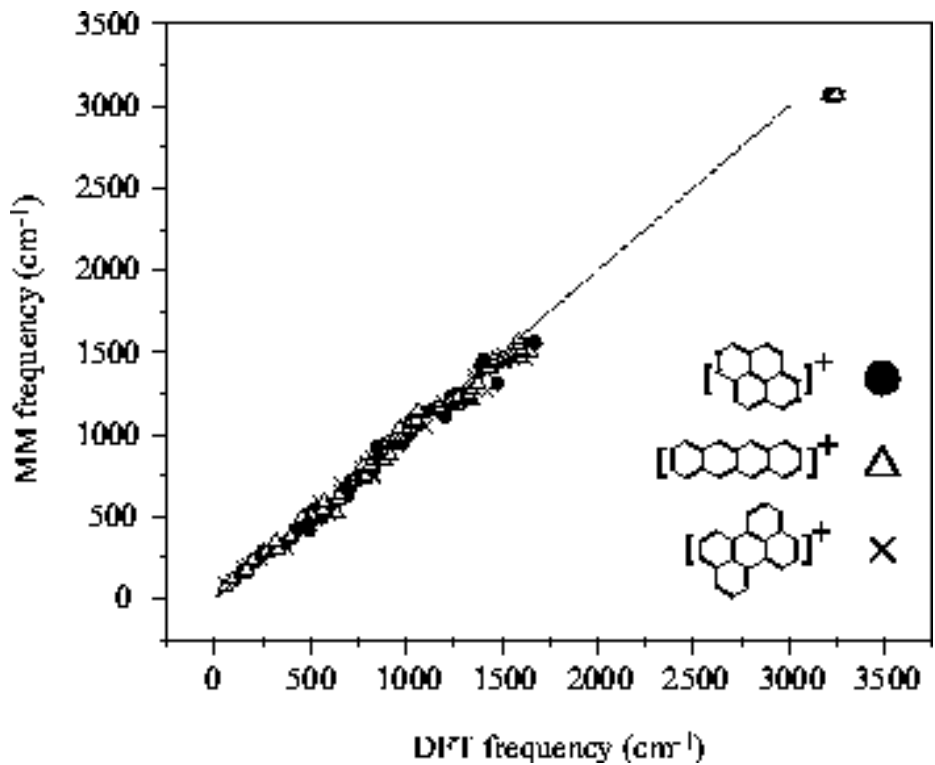}\epsscale{1}
  \caption{\label{dftmm}For molecules larger than coronene DFT
  calculations failed to converge on acceptable ground state
  frequencies and we are forced to resort to a Molecular Mechanical
  model to generate the ground state frequencies.  We confirm the
  validity of the MM model by comparing the resulting ground states to
  those found via DFT for small PAH molecules for where DFT is
  possible.  A one-to-one correspondence is indicated by the dashed
  line.  The large frequencey gap between 1700\,cm$^{-1}$ and
  3100\,cm$^{-1}$ is due to the mass difference between carbon and
  hydrogen, the region around 3100\,cm$^{-1}$ being due to hydrogen
  stretching motions.}
\end{figure}

For molecules the size of coronene and larger, excited state
frequencies were chosen guided by the behaviour of the smaller
systems at the DFT level of theory. Figure \ref{shifts} demonstrates
that most frequency shifts upon excitation fall within an envelope
(shaded). Randomly selecting shifts from this envelope changes the
nature of the sequence structure very little, as shown in figure
\ref{dovetail} for pyrene. Similarly, calculating the ground state
frequencies using molecular mechanics (MM) produces rather similar
results. The correlation between MM and DFT calculated frequencies
is tight, as shown in Figure \ref{dftmm}. To investigate systems
larger than coronene we have invoked a purely molecular mechanical
model, and assumed that these structures behave in the same manner
observed by DFT for the smaller PAH cations. Coronene frequencies
were calculated with both DFT and MM, with both results presented.

\section[]{Molecular mechanical model}
Using parameters taken from the CHARMM force-field
\citep{brooks1983}, vibrational frequencies for electronic ground
states of large PAHs were generated by diagonalization of the
mass-weighted Hessian matrix. Excited state vibrational frequencies
were chosen from the envelope of DFT shifts as outlined above.
Sequence structure was then simulated in an identical manner to the
implicit DFT results.

\subsection{Molecular Mechanical model results}
Using a molecular mechanical (MM) model, we explore a range of larger
PAH molecules, including pyrene, coronene, hexabenzocoronene,
circumcoronene and dicircumcoronene.

\begin{figure}
\plotone{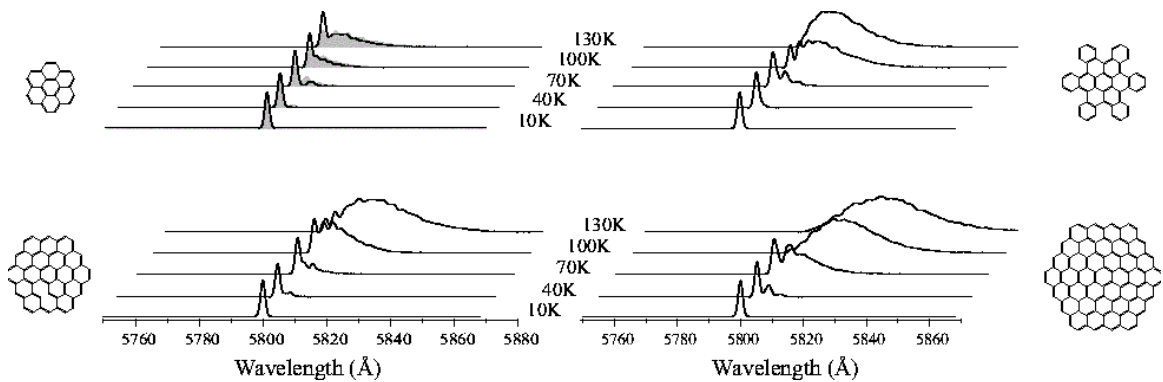}
  \caption{\label{MM sequence}Sequence structure, as a function of
  temperature, is calculated for a number of large PAH molecules using
  the MM model.  For coronene, the largest molecule for which we are
  able to determine reliable frequencies for the ground state via DFT
  , the implicit DFT model (DFT$_i$) is shown by the shaded region.
  Sequence structure, exhibited as a natural consequence of the
  molecular structure of large PAH molecules, closely resembles the
  radial variation in the spectra of the red degraded (RD) Red
  Rectangle bands such as the \rrb\ band.}
\end{figure}

Figure \ref{dovetail} compares the modeled emission spectra of pyrene
at 200\,K from the DFT model and the MM model.  The effect of
increasing the size of the molecule in the MM model is that the
density of states increases very rapidly with energy and thus the
sequence structure is shifted to the red.  Spectra for the coronene,
hexabenzocoronene, circumcoronene and dicircumcoronene at a range of
temperatures are shown in Figure \ref{MM sequence}.

Given the similarity of the sequence structure predicted by the MM
model it is thus difficult to pin-down the exact size of the \rrb\
carrier.

\begin{figure}
\epsscale{.5}\plotone{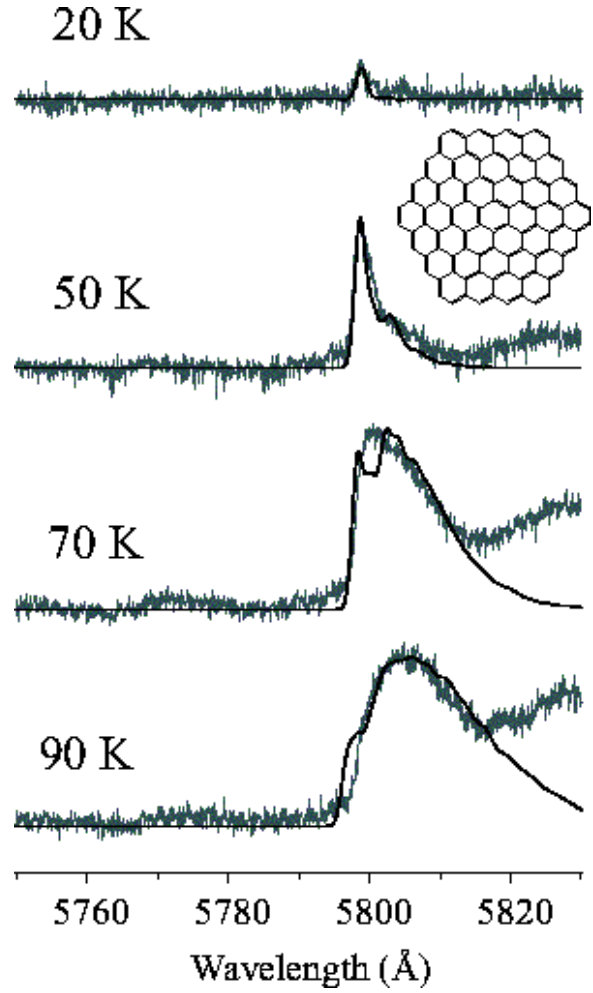}\epsscale{1}
  \caption{\label{fits to spectra} A comparison of the MM spectra, for
  the molecule dicircumcoronene, with the astronomical spectra is
  shown. Choosing dicircumcoronene as our fiducial molecule, the only
  variable in the model is the temperature governing the distribution
  of excited states.  Sequence structure closely mimics the behaviour
  of the observed spectra.  We believe the feature to the right of the
  main band to be a related but largely independent feature of the
  spectrum which is not to be explained by sequence structure from the
  ground state transition (see section \ref{De-hydrogenation}).}
\end{figure}

\begin{figure}
\epsscale{.5}\plotone{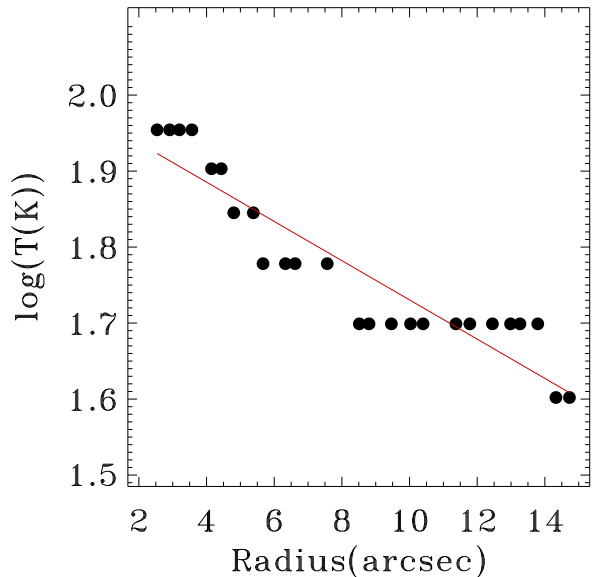}\epsscale{1}
  \caption{\label{temperature sequence} Using the Molecular Mechanical
  model developed in the text and our fiducial molecule,
  dicircumcoronene, one can use the peak shift for the sequence bands
  to estimate a radial temperature variation for the nebula, in this
  instance along arm 2 from Figure \ref{IFU locatation}.}
\end{figure}

Despite the molecules invoked here displaying similar red-shifted
emission spectra, the red-shift of the emission maximum was found to
be very temperature sensitive. As such the MM modeled temperature of
the \rr\ due to sequence emission is relatively insensitive to
carrier(!).  We found that the broadest emission spectra observed in
the \rr\ could be explained in terms of molecules at vibrational
temperatures of $\sim$90\,K. Modeled MM spectra for our fiducial
molecule, dicircumcoronene, are superimposed on \rrb\ RRB in Figure
\ref{fits to spectra}.  An initial estimate of the radial vibrational
temperature of the \rrb\ band carrier is given in Figure
\ref{temperature sequence}.

\subsection{Deficiencies of the MM model}
While the redward shifts in the maxima of emission are reproduced with
little trouble, our model may necessarily over-estimate the size of
the PAH required for the observed (putative) sequence structure. This
is because we select the range of frequency shifts from DFT
calculations necessarily performed on small molecules. One may expect
that larger molecules will exhibit smaller shifts upon excitation as a
consequence of the electronic transition being more delocalized. The
effect of this will be to pull the sequence band contour into the
origin band position, reducing the amount of red shifting. As such,
the observed structure in the RR may be due to a smaller molecule than
dicircumcoronene, at a higher temperature.

Additionally, our model only contains a single temperature population
at this time.  The effect of a compound population along the line of
sight, with a restricted range temperature, would be to smear out the
emission peaks somewhat.

\subsection[]{Should sequence structure be observed?}
\citet{Sarre1995b} note that one might expect sequence structure to be
a feature of molecular emission bands.  If one accepts that large PAH
molecules exist in appreciable quantities in the Red Rectangle, and
that they will be undergoing collisions and photoexcitation, then one
concludes that they will exhibit a significant vibrational
temperature. While many vibrational modes can radiate energy (indeed,
the 3.3$\mu$m and 7.6$\mu$m UIB emission is proposed to originate from
PAH C-H and C-C stretches) there will be many modes which cannot
couple to electromagnetic radiation. As such, vibrational energy can
be ``stored'' in the PAH for an appreciable time. This phenomenon and
the high photon flux in the nebula will bring about a vibrational
temperature. This temperature may exceed 100\,K, and electronically
excited molecules will radiate to vibrational wavefunctions with
nearly identical character to that in the excited state. We conclude
that a large PAH molecule in an environment such as the RR will
inevitably exhibit sequence structure in its emission.

\subsection{De-hydrogenation}
\label{De-hydrogenation}
Kokkin \& Schmidt (in prep.) demonstrate that the effect of
de-hydrogenation on the in-plane electronic transitions in large PAH
molecules is primarily to shift the band a small amount (of the order
of a few $\sim$10\AA) although bond-length alteration induced by
dehydrogenation may be important in systems as large as decacyl.
Since in-plane transitions are the strongest and only involve
molecular orbitals of $\pi$-symmetry ($A''$), dehydrogenation has
little effect on the position of the excitation.  Out-of-plane
transitions of PAH radicals are expected to be significantly shifted
from the strong in-plane transitions of the neutral close-shell
parent, and such spectra will also exhibit a degree of vibrational
structure due to the localization of the lone-pair chromophore.

If the \rrb\ band should indeed be attributed to emission from a 45-100
carbon atom PAH (such as dicircumcoronene) then one may postulate a
second band, at longer wavelengths, derived from the de-hydrogenated
molecule. Such bands would almost certainly be observed in
vibrationally hotter states since one expects these bands to be more
prominent closer to the central star, an energy source capable of
removing the hydrogen atom from the PAH. They would not be observed as
sharp bands such as the RD emission feature. Such bands would better
resemble the \citet{VanWinckel2002} \emph{Symmetric} (S) bands.  The
\rr\ spectrum is replete with such features.  Thus we tentatively
propose the second feature, at $\lambda$5826.5 \citep{VanWinckel2002},
the extreme red limit of our \argus\ spectra, as sequence emission
from a dehydrogenated PAH responsible for the \rrb\ band.

\subsection{Other Red Degraded (RD) bands.}
If one accepts a sequence structure explanation for the \rrb\ RRB, it
is tempting to propose similar explanations for the remaining RD bands
in the \rr. A number of RD bands, each of which can plausibly be
assigned an associated S band derived from the dehydrogenated
molecule, are listed by \citet{VanWinckel2002}. Such bands would not
arise in a common carrier but would instead arise in similar 45-100
carbon atom PAH molecules. Where a PAH has more than one symmetry
unique dehydrogenation site, a single RD band could have associated
with it more than one S band.

\subsection{Symmetry and candidate selection}
When suggesting candidate carriers for the RRBs, one is naturally
tempted to choose highly symmetric molecules in order to avoid the
plurality associated with arbitrary choices. However, there are other
good reasons to suggest a highly symmetric carrier. The
Born-Oppenheimer breakdown which prevents many large molecules from
fluorescing is facilitated by a large background of vibronic states of
the same symmetry. By invoking a more symmetric candidate, the number
of background states of the same symmetry naturally falls, as there
are more irreducible representations of the molecular point group.
As such, it is more likely that a highly symmetric molecule will
fluoresce than a less symmetric one.

\section{Conclusions}
We have demonstrated that the radial variation of the profile of the
\rrb\ \rr\ band is well modeled as molecular sequence structure
arising in a PAH molecule with 45-100 carbon atoms.  Our model has the
attractive properties of :
\begin{itemize}
\item
Not requiring an exotic molecule, once one accepts the presence of
PAHs in the ISM.
\item
Not requiring high nebular temperatures.
\item
Not requiring an exotic emission mechanism.  While sequence
structure does not appear prominently in the astronomical literature,
it is a simple molecular emission phenomenon.
\end{itemize}

If sequence structure is the origin of the \rrb\ RRB then it is
natural to extend the hypothesis to the other Red Degraded RRBs,
invoking sequence structure for additional molecules.

\acknowledgments Acknowledgements : The authors wish to thank ESO
staff for invaluable discussion regarding observing strategy prior to
observation.  We thank the reviewer for insightful comments which led
to a greatly improved manuscript.  We thank ATAC for awarding two
nights of the Australian \flames\ guaranteed time to this program.
RGS thanks ANSTO for financial support to cover travel expenses.  NJR
acknowledges the award of a Gritton Scholarship.  We thank
Prof. P. Thaddeus and Dr. A. Gray-Weale for helpful discussions.
Preparations for the observations reported made extensive use of
excellent archival material from \emph{HST} and the 2MASS and UCAC-2
programs.

\appendix

\section{DFT calculations of band profiles}
\label{dead hard sums}
Assuming no change in vibrational state, the photon energy for
sequence transition $j$, $T_j$, is simply
\begin{equation}
T_j = T_0 + h\sum_{i=1}^{3N-6}n_{ij} (\nu_i' -\nu_i''),
\end{equation}
where $T_0$ is the energy of the origin transition, $N$ is the number
of atoms in the PAH cation and $n_i$ is the number of quanta of
vibrational energy in mode $i$. Simulated emission spectra were
produced by convolving the predicted transitions energies with a
Gaussian profile simulating the rotational structure of each sequence
band.

The vibrational population in the excited electronic state was
assumed to have an exponential dependence on vibrational energy of
the ground vibrational state $j$
\begin{equation}
P_j = \exp(-\beta E''_j),
\end{equation}
where
\begin{equation}
E''_j=h\sum_{i=1}^{3N-6}n_{ij} \nu_i''.
\end{equation}
The simulated emission spectrum $I(\lambda)$ is thus a sum over sets
of quanta (vibrational states)
\begin{equation}
I(\lambda) = \sum_j P_j \exp(-\alpha(\lambda-hc/T_j)^2)
\end{equation}
where $\alpha$ controls the FWHM of the gaussian (rotational)
profile.

In cases where the number of states needed to converge the shape of
the calculated sequence emission profile was prohibitively large, a
Monte Carlo sampling technique was employed. Here, states, $j$, were
chosen as a Monte Carlo Markov Chain and added to the simulation one
at a time (starting at the ground vibrational level). The probability
of accepting a step, in the space of the number of quanta in each
vibrational mode, was given by the Metropolis technique, whereby all
steps down in vibrational energy were chosen with unit probability and
those up in energy were accepted with probability $P_j = \exp(-\beta
E''_j)$. Since the probability of state $j$ being occupied is
naturally taken into the selection algorithm, the emission spectrum
was taken as a simple sum over the sampled states.
\begin{equation}
I(\lambda) = \sum_j \exp(-\alpha(\lambda-hc/T_j)^2).
\end{equation}
Typically, 10000 states were enough to converge the shape of the
emission band profile.

\clearpage
\begin{deluxetable}{ccrlc}
  \tablewidth{0pt}
  \tablecaption{Observation log, 20041229-30, VLT-UT2 ARGUS HR11.\label{observations}}
  \tablehead{\colhead{Pointing} & \colhead{RA/Dec (J2000)} &
  \colhead{Pos. Ang.} & \colhead{Exp. time} & \colhead{Date}}
  \startdata
  P0 & 06:19:58.14  -10:38:14.0 &  12 & 3$\times$2$\times$600sec  & 2005-12-29\\
  P1 & 06:19:58.32  -10:38:23.3 & 165 & 3$\times$2$\times$1800sec & 2005-12-29\\
  P2 & 06:19:57.68  -10:38:20.7 & 230 & 2$\times$3$\times$1200sec & 2005-12-29\\
     &      ' '                 & ' ' & 1$\times$3$\times$1200sec & 2005-12-30\\
  P3 & 06:19:57.95  -10:38:04.5 & 340 & 3$\times$3$\times$1200sec & 2005-12-30\\
  P4 & 06:19:58.63  -10:38:06.4 &  43 & 3$\times$3$\times$1200sec & 2005-12-30\\
   \hline
\enddata
\end{deluxetable}

\clearpage

\begin{thebibliography}{}
\bibitem[\'Ad\'amkovics \etal (2003)]{Adamkovics2003}\'Ad\'amkovics, M., Blake, G.A., McCall, B.J. 2003, \apj, 595, 235
\bibitem[Allamandola \etal (1985)]{Allamandola1985}Allamandola, L.J., Tielens, A.G.G.M., Barker, J.R. 1985, \apj, 290 ,L25
\bibitem[Allamandola \etal (1999)]{Allamandola1999}Allamandola, L.J., Hudgins, D.M., \& Sandford, S.A. 1999, \apjl, 511, L115
\bibitem[Brechignac \& Pino(1999)]{Brechignac1999}Brechignac, P., Pino, T. 1999, \aap, 343, L49
\bibitem[Brooks \etal (1983)]{brooks1983} Brooks, B.R., Bruccoleri, R.E., Olafson, B.D., \etal\ 1983, J. Comp. Chem., 4, 187
\bibitem[Cohen \etal (1975)]{Cohen1975} Cohen, M., Anderson, C.M., Cowley, M., \etal\ 1975, \apj, 196, 179
\bibitem[Cohen \etal (2004)]{Cohen2004} Cohen, M., Winckel, H.V., Bond, W.E., \& Gull, T.R., 2004, \aj, 127, 2362
\bibitem[Ding \etal (2003)]{Ding2003} Ding, H., Schmidt, T.W., Pino, T., Boguslavskiy, A.E., Gathe, F., \& Maier, J.P., 2003, \jcp, 119, 814
\bibitem[Donn, Hodge \& Mentall (1968)]{Donn1968}Donn, B., Hodge, R.C. \& Mentall ,J.E. 1968, \apj, 154, 135
\bibitem[Douglas(1977)]{Douglas1977} Douglas, A.E., 1977, Nature, 269, 130
\bibitem[Fulara \& Krelowski(2000)]{Fulara2000}Fulara, J., \& Krelowski, J., 2000, NewAR, 44, 581
\bibitem[Frisch \etal (2003)]{Frisch2003} Frisch, M.J., Trucks, G.W., Schlegel, H.B., Scuseria, G.E., Robb, M.A., Cheeseman, J.R., Montgomery, J.A. Jr., Vreven, T., \etal\ 2003 \textsc{gaussian 03}, Revision B.05 Gaussian, Inc., Pittsburgh PA
% Kudin, K.N., Burant, J.C., Millam, J.M.,
%S. S. Iyengar, J. Tomasi,
%V. Barone, B. Mennucci, M. Cossi, G. Scalmani, N. Rega,
%G. A. Petersson, H. Nakatsuji, M. Hada, M. Ehara, K. Toyota,
%R. Fukuda, J. Hasegawa, M. Ishida, T. Nakajima, Y. Honda, O. Kitao,
%H. Nakai, M. Klene, X. Li, J. E. Knox, H. P. Hratchian, J. B. Cross,
%C. Adamo, J. Jaramillo, R. Gomperts, R. E. Stratmann, O. Yazyev,
%A. J. Austin, R. Cammi, C. Pomelli, J. W. Ochterski, P. Y. Ayala,
%K. Morokuma, G. A. Voth, P. Salvador, J. J. Dannenberg,
%V. G. Zakrzewski, S. Dapprich, A. D. Daniels, M. C. Strain,
%O. Farkas, D. K. Malick, A. D. Rabuck, K. Raghavachari,
%J. B. Foresman, J. V. Ortiz, Q. Cui, A. G. Baboul, S. Clifford,
%J. Cioslowski, B. B. Stefanov, G. Liu, A. Liashenko, P. Piskorz,
%I. Komaromi, R. L. Martin, D. J. Fox, T. Keith, M. A. Al-Laham,
%C. Y. Peng, A. Nanayakkara, M. Challacombe, P. M. W. Gill,
%B. Johnson, W. Chen, M. W. Wong, C. Gonzalez, and J. A. Pople,
%Gaussian, Inc., Pittsburgh PA, 2003.
\bibitem[Geballe, Tielens, Allamandola \etal\ (1989)]{Geballe1989} Geballe, T.R., Tielens, A.G.G.M., Allamandola, L.J., Moorhouse, A. \& Brand, P.W.J.L., 1989, \apj, 341, 278
\bibitem[Glinski \& Anderson(2002)]{Glinski2002}Glinski, R.J., \& Anderson, C.M. 2002, \mnras, 332, 17
\bibitem[Glinski \& Nuth(1997)]{Glinski1997} Glinski, R.J., \& Nuth, J.A. \textsc{iii} 1997, Ap\&SS, 249, 143
\bibitem[Gillingham \etal (2003)]{Gillingham2003}Gillingham, P.R., Popovic, D., Farrell, T.J., Waller, L.G. 2003, SPIE, 4841, 1170
\bibitem[Heinecke(1998)]{Heinecke1998}Heinecke, E., Hartmann, D., Muller, R., \& Hese, A. 1998, J. Chem. Phys., 109, 906
\bibitem[Herbig(1995)]{Herbig1995}Herbig, G.H.  1995, \araa, 33, 1
\bibitem[Kerr \etal (1998)]{Kerr1998}Kerr, T.H., Hibbins, R.E., Fossey, S.J., Miles, J.R., Sarre, P.J. 1998, \apj, 495, 941
\bibitem[Kroto \etal (1985)]{Kroto1985} Kroto, H.W., Heath, J.R., O'Brien, S.Co, Curl, R.F., \& Smalley, R.E. 1985, Nature, 318, 14
\bibitem[L\'eger \& Dhendecourt (1985)]{Leger1985}L\'eger, A. \& Dhendecourt, L. 1985, \aap, 146, 81L
\bibitem[Linnartz \etal (2004)]{Linnartz2004} Linnartz, H., Araki, M., Ding, H., Boguslavskiy, A.E., Kolek, P., Schmidt, T.W., Motylewski, T., Cias, P. \etal\ 2004, \apj, 616, 1301
\bibitem[Maier \etal (2001)]{Maier2001}Maier, J.P., Lakin, N.M., Walker, G.A.H., Bohlender, D.A. 2001, \apj, 553, 267
\bibitem[Men'shchikov \etal (2002)]{Men'shchikov2002}Men'shchikov, A.B, Schertl, D. Tuthill, P.G., Weigelt, G., \& Yungelson, L.R. 2002, \aap, 393, 867
\bibitem[Merrill(1934)]{Merrill1934}Merrill, P.W. 1934, Publ. Astron.Soc Pac., 46, 206
\bibitem[Merrill(1936)]{Merrill1936}Merrill, P.W. 1936, Publ. Astron.Soc Pac., 48, 179
\bibitem[Merrill \& Wilson(1938)]{Merrill1938}Merrill, P.W., \& Wilson O.C. 1938, \apj, 87, 9
\bibitem[Miyata \etal (2004)]{Miyata2004}Miyata, T., Kataza, H., Okamoto, Y.K. \etal\ 2004, \aap, 415, 179
\bibitem[Peeters \etal (2004)]{Peeters2004} Peeters, E., Spoon, H.W.W., \& Tielens, A.G.G.M.  2004, \apj, 613, 986
\bibitem[Press \etal (1993)]{numrepc} Press, W.H., \etal\ 1993, Numerical Recipes in C : The Art of Scientific Computing (2nd ed.; Cambridge University Press)
\bibitem[Rouan \etal (1997)]{Rouan1997} Rouan, D., L\'eger, A., \& Le Coupanec, P. 1997, A\&A ,324, 661
\bibitem[Sarre \etal (1995a)]{Sarre1995a}Sarre, P.J., Miles, J.R., Kerr, T.H., Hibbins, R.E., Fossey, S.J., \& Somerville, W.B. 1995, \mnras, 277, L41
\bibitem[Sarre \etal (1995b)]{Sarre1995b}  Sarre, P.J., Miles, J.R., Scarrott, S.M. 1995, Sci, 269, 674
\bibitem[Scarrott \etal (1992)]{Scarrott1992}Scarrott, S.M., Watkin, S., Miles, J.R. \& Sarre, P.J. 1992, \mnras, 255, 11
\bibitem[Schmidt \etal (2003)]{Schmidt2003a}Schmidt, T.W., Boguslavskiy, A.E., Ding, H., \etal\ 2003, International Journal of Mass Spectrometry, 228, 647
\bibitem[Schmidt \etal (1980)]{Schmidt1980}Schmidt, G.D., Cohen, M., \& Margon, B. 1980, \apjl, 239, L133
\bibitem[Schmidt \& Sharp(2005)]{Schmidt2005}Schmidt, T.W., \& Sharp, R.G. 2005, Aust. J. Chem., 58N2, 69 (arXiv:astro-ph/0501180)
\bibitem[Thaddeus \& McCarthy(2001)]{Thaddeus2001}Thaddeus, P., \& McCarthy, M.C. 2001, Spectrochimica, 57, 757
\bibitem[Thorburn \etal (2003)]{Thorburn2003}Thorburn, J.A., Hobbs, L.M., McCall, B.J., Oka, T., Welty, D.E., Friedman, S.D., Snow, T.P., Sonnentrucker, P., \& York, D.G. 2003, \apj, 584, 339
\bibitem[Van Winckel \etal (2002)]{VanWinckel2002}Van Winckel, H., Cohen, M., \& Gull, T.R. 2002, \aap, 390, 147
\bibitem[van der Zwet \&  Allamandola (1985)]{Zwet1985}van der Zwet, G.P., \&  Allamandola, L.J., 1985, \aap, 146, 76
\end{thebibliography}
\end{document}